# Option pricing for Informed Traders

Stoyan V. Stoyanov
Stony Brook University
Email: stoyan.stoyanov@stonybrook.edu

Yong Shin Kim
Stony Brook University
Email: aaron.kim@stonybrook.edu

Svetlozar T. Rachev
Stony Brook University
Email: svetlozar.rachev@stonybrook.edu

Frank J. Fabozzi
EDHEC Business School
Email: fabozzi321@aol.com

ABSTRACT

In this paper we extend the theory of option pricing to take into account and explain the empirical evidence for asset prices such as non-Gaussian returns, long-range dependence, volatility clustering, non-Gaussian copula dependence, as well as theoretical issues such as asymmetric information and the presence of limited arbitrage opportunities

Corresponding Author: Frank J. Fabozzi
EDHEC Business School
Email: fabozzi321@aol.com

**I. Introduction**

The theory of option pricing (T0P), developed in the seminal works of Black and Scholes (1973) and Merton (1973), provides the theory of finance with the fundamentals to understand, model, estimate and apply the processes by which financial assets are priced. Several works provide a comprehensive exposition of TOP such as Cochrane (2001), Dudley (2001), Skiadas (2009), Campbell (2000, Çelik Ş. (2012), and Munk (2013). Although it is impossible to overlook TOP's enormous influence on the theory of finance and its applications, there are some limitations of the original formulation of TOP coming from some restrictive premises of the theory that are inconsistent with the findings of empirical studies on asset pricing processes. Those phenomena are:

- *Phenomenon 1*. (Empirical evidence). Long-range dependence in asset price time series, volatility clustering of asset returns, skewness of the distribution of asset returns, heavy tails of the distribution of asset returns, multivariate tail dependencies in the vector of asset returns[1];
- *Phenomenon 2*. Market participants often deal with asymmetric information[2],
- *Phenomenon 3*. Prices are often predictable[3]
- *Phenomenon 4*. Markets exhibit limited arbitrage opportunities[4]
- *Phenomena 5*. Markets exhibit chaotic behavior (often referred to as irrational[5]).[6]

Our paper attempts to extend the boundaries of TOP to address those issues and is organized as follows. Section 2 introduces the T)P for informed traders. Section 3 introduces the TAP in the presence of limited arbitrage.

## 2. TOP FOR INFORMED TRADERS

There is a vast amount of literature on asset pricing with asymmetric information and arbitrage opportunities, most notably the models proposed by both Kyle (1985) and Back (1992). Both models assume a market with a continuous-time risky asset and asymmetric information. In the Kyle model there are three financial agents: the market maker, an insider trader (who knows a payoff which will be revealed at a pre-specified future time), and an uninformed (noisy) trader. The market maker has to define a pricing rule in such a way that an equilibrium exists between the traders. Back (1993) extended the model to continuous time.[7] A second line of research concerns studies of markets with asymmetric information based on an enlargement of the filtration and the change of the probability measure, the study by Aase et al. (2010) being one example. Our work is of a different nature. We derive option pricing formulas when traders could have additional information about future asset prices. The trader's information is multifaceted and any general definition will be restrictive in view of the trader's particular trading activities. We derive the analog of the Black-Scholes formula for informed traders, and construct and estimate an implied information surface based on our option pricing formula

---

[1] See Rachev and Mittnik (2000), Schoutens (2003), Cont and Takov (2004), Rachev et al. (2011). and the references therein.
[2] See Brunnermeier (2001) and Kelly and Ljungqvist (2008).
[3] See Campbell and Yogo (2005), Boucher (2006), Ang and Bekaert (2007), and Caporin et al. (2013)
[4] See Lo (1991), Campbell, et al.(1997), Andersson(1998), Diebold (2001). Nielsen (2010), Johansen (2011), and Caporale and Gil-Alana (2014).
[5] The earliest known evidence for this saying appeared in a column by Gary Shilling in *Forbes* magazine in February 1993 (http://quoteinvestigator.com/2011/08/09/remain-solvent/).
[6] See Hsieh (1990), Trippi (1995), Banerjee S. (2013), Jovanovic and Schinckus (2013), Rubinstein (2001), Shiller (2003), Daniel and Titman (1999), Pedersen (2015), Harford and Alexander (2013), Bloxham (2016), and Farmer (2014).

## 2.1. Option Pricing Formula for a Trader with Prior Information on the Stock-Price Direction

In this section we introduce the notion of an informed trader (designated as $\aleph_\tau, \tau \in R$) in order to extend the Cox-Ingersoll-Ross (CRR) binomial pricing model (Cox et al.,1979) for informed (resp. misinformed) traders $\aleph_\tau, \tau > 0$ (resp. $\aleph_\tau, \tau < 0$). As will be seen, the CRR-model is a particular case of our model when the trader has no additional information (i.e., $\tau = 0$). $\aleph'_\tau$s information manifests in a superior trading performance of $\aleph_\tau$ with $\tau > 0$ over an uninformed $\aleph_0$-trader. Indeed, there is no way to have a unique "ideal" definition of "information level" $\tau$. The information level will depend on the nature of trading activities that $\aleph_\tau, \tau > 0$ is involved in. We start our analysis recalling the CRR binomial model: at $t$ the stock price is $x = S_t$ and the price in the next discrete period of time is given by

$$S_{t+\Delta t} = \begin{cases} S_{t+\Delta t;up} := xu_{\Delta t} = xe^{\sigma\sqrt{\Delta t}} \text{ w. p. } p_{\Delta t} = \frac{1}{2} + \frac{\mu - \frac{\sigma^2}{2}}{2\sigma}\sqrt{\Delta t} \\ S_{t+\Delta t;down} := xd_{\Delta t} = xe^{-\sigma\sqrt{\Delta t}} \text{ w. p. } 1 - p_{\Delta t} \end{cases} \quad (1)$$

A European contingent claim ECC pays $f_{t+\Delta t} = f_{t+\Delta t;up}$ (resp. $f_{t+\Delta t;down}$) with probability $p_{\Delta t}$ (resp. $1 - p_{\Delta t}$). Trader $\aleph_\tau$ ($\tau > 0$) takes a short position in the EEC. Suppose at $t$, $\aleph_\tau$ ($\tau > 0$) knows the price direction at $t + \Delta t$ with probability $p_\tau \in (0,1)$. Because the natural probability $p_{\Delta t}$ is now assumed fixed (cannot be conveniently chosen), we need the following extension of the CRR-model [8] and the Jarrow-Rudd model [9]. To this end, we have the following lemma due to Kim et al. (2016) (referred to hereafter as KSRF). The corresponding tree's weak convergence (denoted as " $\Rightarrow^w$ "), shown in KSRF, is based on Proposition 3 in Davydov and Rotar (2008).

*Lemma 1. Let the stock-price process dynamics on natural world $\mathbb{P}$ is given by geometric Brownian motion (GBM)*

$$S_{t;\mathbb{P}} = S_{t;\mathbb{P}}^{(\mu,\sigma)} = e^{\left(\mu - \frac{1}{2}\sigma^2\right)t + \sigma B(t)}, t \geq 0 \quad (2)$$

---

[8] In CRR-model $p_{\Delta t}$ is determined in (1.1), its estimation is based on estimates of $\mu$ and $\sigma^2$, while in our approach, we suggest that $p_{\Delta t}$ is estimated separately from the $\mu$ and $\sigma^2$, using the time series of signs of $\Delta t$ −returns. This makes our model more flexible in the discrete binomial setting. As far as the limiting continuous price process is concerned, our and the CRR-model indeed coincide, see Lemma 1 next.

[9] In the model of Jarrow and Rudd (1983) the risk-neutral probability $q_{\Delta t}$ is chosen to be ½, in our model $q_{\Delta t} := p_{\Delta t} - \sqrt{p_{\Delta t}(1 - p_{\Delta t})}\theta\sqrt{\Delta t}$ will depend on the physical measure $p_{\Delta t}$ (resolving the "discontinuous puzzle", see Kim et al. (2016) ). Indeed, as $\Delta t \downarrow 0$, our and Jarrow-Rudd model converge to the same risk neutral GBM, see Lemma (1) next.

*and on the risk-neutral world $\mathbb{Q}$ is given by*

$$S_{t;\mathbb{Q}} = S_{t;\mathbb{Q}}^{(r,\sigma)} = e^{\left(r-\frac{1}{2}\sigma^2\right)t + \sigma B^{\mathbb{Q}}(t)}, t \geq 0 \tag{3}$$

*where $B^{\mathbb{Q}}$ is Brownian motion on $\mathbb{Q}$ and an arithmetic Brownian motion (ABM) on $\mathbb{P}$, $B^{\mathbb{Q}}(t) = B(t) + \theta t$ and $\theta = \frac{\mu - r}{\sigma}$ is the market price of risk. Fix $p \in (0,1)$. Then the binomial tree approximating $S_{t;\mathbb{P}}, t \in [0, T]$ ( in the sense of " $\Rightarrow^w$ ") with given probability for upward stock movement $p_{\Delta t}$ is given by*

$$S_{(k+1)\Delta t;\mathbb{P}} := S_{k\Delta t;\mathbb{P}} \begin{cases} 1 + \mu\Delta t + \sqrt{\frac{1-p_{\Delta t}}{p_{\Delta t}}} \sigma\sqrt{\Delta t} \ w.p. \ p_{\Delta t}, \\ 1 + \mu\Delta t - \sqrt{\frac{p_{\Delta t}}{1-p_{\Delta t}}} \sigma\sqrt{\Delta t} \ w.p. \ 1 - p_{\Delta t} \end{cases} \tag{4}$$

$k := 0, \dots, n-1, n\Delta t = T$. *The corresponding risk-neutral tree approximating $S_{t;\mathbb{Q}}$ is given by*

$$S_{(k+1)\Delta t;\mathbb{Q}} := S_{k\Delta t;\mathbb{Q}} \begin{cases} 1 + \mu\Delta t + \sqrt{\frac{1-p_{\Delta t}}{p_{\Delta t}}} \sigma\sqrt{\Delta t} \ w.p. \ q_{\Delta t}, \\ 1 + \mu\Delta t - \sqrt{\frac{p_{\Delta t}}{1-p_{\Delta t}}} \sigma\sqrt{\Delta t} \ w.p. \ 1 - q_{\Delta t} \end{cases} \tag{5}$$

*where $q_{\Delta t} := p_{\Delta t} - \sqrt{p_{\Delta t}(1-p_{\Delta t})}\, \theta\sqrt{\Delta t}$ is the risk-neutral probability for an upward stock movement.*

Suppose at each instant $t$ ($t = k\Delta t, k = 0, \dots, N-1$), $N\Delta t = T$), $\aleph_\tau$ makes an independent informed prediction that the stock will be "up" at $t + \Delta t$ with probability $p_{\tau,t}$ for correct prediction. Because we assume that the probability $p_{\tau,t} = p_\tau$ is constant, we are dealing with a Bernoulli sequence of trader's guesses about the stock direction. If at t, $\aleph_\tau$ believes that the stock will be "up" at $t + \Delta t$, that is $u_{\Delta t} = x\left(1 + \mu\Delta t + \sqrt{\frac{1-p_{\Delta t}}{p_{\Delta t}}}\sigma\sqrt{\Delta t}\right)$, $x = S_t$, (resp. "down", that is, $d_{\Delta t} = x\left(1 + \mu\Delta t - \sqrt{\frac{p_{\Delta t}}{1-p_{\Delta t}}}\sigma\sqrt{\Delta t}\right)$) at $t + \Delta t$, the informed trader enters a long-forward[10] (resp. short-forward) contract with an un-informed $\aleph_0$ −trader [11]

---

[10] The forward price is $xe^{r\Delta t} = x(1 + r\Delta t)$ with the contract's maturity $\Delta t$.
[11] Applying the tree model from Lemma 1, $\aleph'_\tau s$ payoff (at time $t + \Delta t$,) is

$$P_{t+\Delta t}(\aleph_\tau) = x \begin{cases} u_{\Delta t} - e^{r\Delta t} \ w.p. \ p_{\Delta t} p_{\tau,t} \\ e^{r\Delta t} - u_{\Delta t} \ w.p. \ (1 - p_{\tau,t})p_{\Delta t} \\ e^{r\Delta t} - d_{\Delta t} w.p. \ p_{\tau,t}(1 - p_{\Delta t}) \\ d_{\Delta t} - e^{r\Delta t} \ w.p. \ (1 - p_{\tau,t})(1 - p_{\Delta t}) \end{cases}$$

Suppose $\aleph_\tau$ takes a short position in a European option contract with payoff $f_{u,\Delta t}$ (resp. $f_{d,\Delta t}$) when the stock-price is up (resp. down), at maturity $t + \Delta t$. To hedge the risk of the informed trader's position $\aleph_\tau$ takes (i)$\Delta^{(\tau)}$ −position in the underlying stock to hedge the "up" and "down" risk; (ii) $\Delta_{\tau;u}$(resp. $\Delta_{\tau;d}$) − position in arbitrage shares to hedge the "information" risk, when $\aleph_\tau$ believes that the stock will be "up" (resp. "down"). Equating all outcomes of

Page | 3

We now formulate the first version of the Black-Scholes formula for an informed trader $\aleph_\tau, \tau \in R$, with some knowledge of stock direction (we treat the case of $\aleph_\tau, \tau < 0$, as dealing with misinformed trader

*Proposition 1. Suppose the trader $\aleph_\tau, \tau \in R$ has information about the instantaneous stock-price direction leading to a specific for $\aleph_\tau$-risk- neutral dynamics given by $dS_t^{(\aleph_\tau)} = (r - C_\tau \tau)S_t^{(\aleph_\tau)}dt + \sigma S_t^{(\aleph_\tau)}dB(t)$. Then the $\aleph_\tau$'s value of the trader's short position in a European call contract with maturity price $C^{(\aleph_\tau)}\left(S_T^{(\aleph_\tau)}, T\right) = max\left(0, S_T^{(\aleph_\tau)} - K\right)$ is given by*

$$C^{(\aleph_\tau)}\left(S_t^{(\aleph_\tau)}, t\right) = e^{-I^{(\aleph_\tau)}(T-t)}S_t^{(\aleph_\tau)}N(d_1) - Ke^{-r(T-t)}N(d_2) \qquad (1.6)$$

*where* $I^{(\aleph_\tau)} = C_\tau \tau, \tau \in R, C_\tau \geq 0$ $d_1 = \dfrac{\ln\left(\dfrac{S_t^{(\aleph_\tau)}}{K}\right) + \left(r - I^{(\aleph_\tau)} + \dfrac{\sigma^2}{2}\right)(T-t)}{\sigma\sqrt{T-t}}$ *and* $d_2 = d_1 - \sigma\sqrt{T-t}$.

## 2.2. Option Pricing Formula for a Trader with Information On The Mean Stock-Return

Using the same setting as in Section 1.1, suppose the true price dynamics is given by $S_{t;\mathbb{P}}^{(\mu,\sigma)}$ by (2). However, here the stock is actually traded with market perceived dynamics $S_{t;\mathbb{P}}^{(\nu,\sigma)} = e^{\left(\nu - \frac{1}{2}\sigma^2\right)t + \sigma B(t)}, t \geq 0$, for some $\nu > r > 0$. Suppose at each instant $t = k\Delta t, k = 0, \ldots, N-1$, $\aleph_\tau$ knows whether the true mean return $\mu$ is above or below the market value of $\nu$ with success probability $p_{\tau,t} = p_\tau$. [12]

In the next lemma we fix the mean drift and determine the corresponding trees approximating (2) and (3).

---

the riskless portfolio and applying Lemma 1 leads to
$\Delta_{\tau;d} = -\Delta_{\tau;u}, \Delta^{(\tau)} = \Delta^{(0)} + +\Delta_{\tau;u}(2 q_{\Delta t} - 1), \Delta^{(0)} = \dfrac{f_{u,\Delta t} - f_{d,\Delta t}}{u_{\Delta t} - d_{\Delta t}}$.
Setting $\Delta_{\tau;u} = -C_\tau \dfrac{\tau}{\sigma}\Delta^{(0)}\sqrt{\Delta t}$ and discounting the instantaneous riskless portfolio delivers the option value at
$t: f_t = e^{-r\Delta t}\left(q^{(\aleph_\tau)}f_{u,\Delta t} + \left(1 - q^{(\aleph_\tau)}\right)f_{d,\Delta t}\right)$.
The discount factor $q^{(\aleph_\tau)}$ has two equivalent forms:

$$q^{(\aleph_\tau)} := q_{\Delta t} - 2C_\tau \dfrac{\tau}{\sigma}\sqrt{\Delta t}q_{\Delta t}(1 - q_{\Delta t}) = \dfrac{1}{2} + \dfrac{r - C_\tau \tau - \dfrac{\sigma^2}{2}}{2\sigma}\sqrt{\Delta t} \;,$$

where $C_\tau \geq 0$ is the coefficient for $\aleph_\tau$'s information level ($\tau < 0$ designates the level of misinformation ). Now, $q^{(\aleph_\tau)}$ converges to 1 (resp. to 0) when the natural probability $p_{\Delta t}$ converges to 1 (resp. 0) . This resolves the discontinuity problem in option pricing(see KSRF). Furthermore, the continuous dynamics of the $\aleph_\tau$'s risk-neutral price process is given by GBM

$$dS_t^{(\aleph_\tau)} = (r - C_\tau \tau)S_t^{(\aleph_\tau)}dt + \sigma S_t^{(\aleph_\tau)}dB(t), \text{ where } C_\tau \tau \in R$$

is designating the level of $\aleph_\tau$'s "clairvoyance". Also assuming that $\Delta_{\tau;u}$ follows its own binomial tree independent of that for $S_t^{(\aleph_\tau)}$ will lead to a subordinated-in-the-mean pricing model for $S_t^{(\aleph_\tau)}$).

[12] For simplicity of exposition, we assume that $p_\tau = 1$. In the general case $p_\tau \in [0,1]$ a more complex tree, similar to the one to be discussed in Section 3, should be used.

*Lemma 2.* Let (2) hold and $v > r > 0$ is fixed. Then the binomial tree

$$S_{(k+1)\Delta t; \mathbb{P}} := S_{k\Delta t; \mathbb{P}} \begin{cases} 1 + v\Delta t + \sigma\sqrt{\Delta t} & w.p. \ p_{v,\Delta t} = \frac{1}{2} + \frac{\mu - v}{2\sigma}\sqrt{\Delta t}, \\ 1 + v\Delta t - \sigma\sqrt{\Delta t} & w.p. \ 1 - p_{v,\Delta t} \end{cases}$$

$k := 0, \ldots, n-1, n\Delta t = T$ is approximating $S_{t;\mathbb{P}}, t \in [0, T]$ in $\Rightarrow^w$-sense. The corresponding risk-neutral tree approximating $S_{t;\mathbb{Q}}$ is given by

$$S_{(k+1)\Delta t; \mathbb{Q}} := S_{k\Delta t; \mathbb{Q}} \begin{cases} 1 + v\Delta t + \sigma\sqrt{\Delta t} & w.p. \ q_{v,\Delta t}, \\ 1 + v\Delta t - \sigma\sqrt{\Delta t} & w.p. \ 1 - q_{v,\Delta t} \end{cases}$$

where $q_{v,\Delta t} = \frac{1}{2} + \frac{r-v}{2\sigma}\sqrt{\Delta t}$ is the risk-neutral probability for upward stock movement.[13]

If $\aleph_\tau$ believes that $\mu > v$ (resp. $\mu < v$), then the informed investor enters a long-forward (resp. short-forward) contract with an un-informed $\aleph_0$−trader. Suppose $\aleph_\tau$ takes a short position in a European option contract with payoff $f_{u,\Delta t}$ (resp. $f_{d,\Delta t}$) when the stock-price is up (resp. down) at maturity $t + \Delta t$[14].

*Proposition 2.* Suppose trader $\aleph_\tau, \tau \in R$ has information about the stock-price mean return leading to a specific for $\aleph_\tau$-risk neutral dynamics given by $dS_t^{(\aleph_\tau)} = \left(r - J_v^{(\aleph_\tau)}\right)S_t^{(\aleph_\tau)}dt + \sigma S_t^{(\aleph_\tau)}dB(t)$. Then the $\aleph_\tau$'s value of the trader's short position in an European call contract with maturity price $C^{(\aleph_\tau)}\left(S_T^{(\aleph_\tau)}, T\right) = \max\left(0, S_T^{(\aleph_\tau)} - K\right)$ is given by

$$S_{(k+1)\Delta t; \mathbb{P}} := S_{k\Delta t; \mathbb{P}} \begin{cases} 1 + \mu\Delta t + \sigma\sqrt{\Delta t} & w.p. \ p = \frac{1}{2}, \\ 1 + \mu\Delta t - \sigma\sqrt{\Delta t} & w.p. \ 1 - p \end{cases}$$

and the second one on the risk-neutral world,

$$S_{(k+1)\Delta t; \mathbb{Q}} := S_{k\Delta t; \mathbb{Q}} \begin{cases} 1 + v\Delta t + \sigma\sqrt{\Delta t} & w.p. \ q = \frac{1}{2} + \frac{\mu - v}{2\sigma}\sqrt{\Delta t} \\ 1 + v\Delta t - \sigma\sqrt{\Delta t} & w.p. \ 1 - q \end{cases}$$

---

[13] Lemma 2 could be viewed as a "discrete version of the Girsanov's formula" meaning that we have binomial approximations for $S_{t;\mathbb{P}} = S_{t;\mathbb{P}}^{(\mu,\sigma)} = e^{\left(\mu - \frac{1}{2}\sigma^2\right)t + \sigma B(t)}, t \geq 0$ and $S_{t;\mathbb{Q}} = S_{t;\mathbb{Q}}^{(r,\sigma)}$: the first one on the natural world

[14] From Lemma it follows that the value of the option contract at t is $f_t = e^{-r\Delta t}\left(q_v^{(\aleph_\tau)}f_{u,\Delta t} + \left(1 - q_v^{(\aleph_\tau)}\right)f_{d,\Delta t}\right)$,

where $q_v^{(\aleph_\tau)} = q_{v,\Delta t} - 2C_\tau \frac{\tau}{\sigma}\sqrt{\Delta t}q_{v,\Delta t}(1 - q_{v,\Delta t}) = \frac{1}{2} + \frac{1}{2} + \frac{r - J_v^{(\aleph_\tau)} - \frac{\sigma^2}{2}}{2\sigma}\sqrt{\Delta t}$ and $J_v^{(\aleph_\tau)} := \left(\frac{1}{2}\right)(C_\tau \tau v - \sigma^2)$. Now the continuous dynamics of the $\aleph_\tau'$s risk-neutral price process is given by the GBM $dS_t^{(\aleph_\tau)} = \left(r - J_v^{(\aleph_\tau)}\right)S_t^{(\aleph_\tau)}dt + \sigma S_t^{(\aleph_\tau)}dB(t)$, where the yield $J_v^{(\aleph_\tau)}$ now depends on the market perceived mean stock-price return ($v$) and volatility ($\sigma$).

Page | 5

$$C^{(\aleph_\tau)}\left(S_t^{(\aleph_\tau)}, t\right) = e^{-J_v^{(\aleph_\tau)}(T-t)} S_t^{(\aleph_\tau)} N(d_1) - K e^{-r(T-t)} N(d_2) \tag{7}$$

where $J_v^{(\aleph_\tau)} = \left(\frac{1}{2}\right)(C_\tau \tau v - \sigma^2), C_\tau \geq 0$ $d_1 = \dfrac{\ln\left(\frac{S_t^{(\aleph_\tau)}}{K}\right) + \left(r - J_v^{(\aleph_\tau)} + \frac{\sigma^2}{2}\right)(T-t)}{\sigma\sqrt{T-t}}$ and $d_2 = d_1 - \sigma\sqrt{T-t}$.

### 2.3. Option Pricing Formula For a Trader With Information On The Stock-Return Mean and Volatility

We use the same setting as in Section 2.2 but now we assume that $\aleph_\tau$ knows that the true stock-price dynamics is given by $S_{t;\mathbb{P}}^{(\mu,\sigma)}$ in (2) with Sharpe ratio $\theta = \frac{\mu-r}{\sigma}$. The stock is, however, traded with market perceived dynamics $S_{t;\mathbb{P}}^{(\gamma,\varrho)} = e^{\left(\gamma - \frac{1}{2}\varrho^2\right)t + \varrho B(t)}, t \geq 0$ and market perceived Sharpe ratio $\phi = \frac{\gamma - r}{\varrho}$. To explain the arbitrage strategy $\aleph_\tau$ ($\tau > 0$) will employ, we need the following analog of Lemmas 1 and 2, the proof based on the techniques developed in KSRF.

*Lemma 3. Let (2) hold. Fix the mean return $\gamma > r > 0$ and the volatility level $\varrho > 0$. Then the trinomial tree approximating $S_{t;\mathbb{P}}, t \in [0,T]$ in the sense of $\Rightarrow^w$ is given by*

$$\frac{S_{(k+1)\Delta t;\mathbb{P}}}{S_{k\Delta t;\mathbb{P}}} = 1 + \gamma \Delta t + \begin{cases} \varrho\sqrt{\Delta t} & w.p. \frac{\sigma^2}{2\varrho^2} + (\theta - \phi)\sqrt{\Delta t} \\ 0 & w.p. 1 - \frac{\sigma^2}{\varrho^2} \\ -\varrho\sqrt{\Delta t} & w.p. \frac{\sigma^2}{2\varrho^2} - (\theta - \phi)\sqrt{\Delta t} \end{cases} \tag{8}$$

*The corresponding risk-neutral tree approximating $S_{t;\mathbb{Q}}$ is given by[15]*

---

[15] Lemma 3 could be viewed as a discrete extension of Girsanov's formula when the volatility has been changed together with the mean drift. Indeed, we have trinomial approximations for $S_{t;\mathbb{P}} = S_{t;\mathbb{P}}^{(\mu,\sigma)}$ and $S_{t;\mathbb{Q}} = S_{t;\mathbb{Q}}^{(r,\sigma)}$

$S_{t;\mathbb{P}} = S_{t;\mathbb{P}}^{(\mu,\sigma)} = e^{\left(\mu - \frac{1}{2}\sigma^2\right)t + \sigma B(t)}, t \geq 0$. Namely, $\frac{S_{(k+1)\Delta t;\mathbb{P}}}{S_{k\Delta t;\mathbb{P}}} = 1 + \mu\Delta t + \begin{cases} \sigma\sqrt{\Delta t} & w.p. \frac{1}{2} \\ 0 & w.p. 0 \\ -\sigma\sqrt{\Delta t} & w.p. \frac{1}{2} \end{cases}$ and $\frac{S_{(k+1)\Delta t;\mathbb{Q}}}{S_{k\Delta t;\mathbb{Q}}} = 1 +$

$\gamma\Delta t + \begin{cases} \varrho\sqrt{\Delta t} & w.p. \frac{\sigma^2}{2\varrho^2} + \frac{\mu-\nu}{\sigma}\sqrt{\Delta t} \\ 0 & w.p. 1 - \frac{\sigma^2}{\varrho^2} \\ -\varrho\sqrt{\Delta t} & w.p. \frac{\sigma^2}{2\varrho^2} - \frac{\mu-\nu}{\sigma}\sqrt{\Delta t} \end{cases}$.

$$\frac{S_{(k+1)\Delta t;\mathbb{Q}}}{S_{k\Delta t;\mathbb{Q}}} := 1 + \gamma\Delta t + \begin{cases} \varrho\sqrt{\Delta t} & w.p.\ \frac{\sigma^2}{2\varrho^2} - \phi\sqrt{\Delta t} \\ 0 & w.p.\ 1 - \frac{\sigma^2}{\varrho^2} \\ -\varrho\sqrt{\Delta t} & w.p.\ \frac{\sigma^2}{2\varrho^2} + \phi\sqrt{\Delta t} \end{cases} \qquad (9)$$

Suppose $\aleph_\tau$ knows the market direction with probability 1 but the trader is permitted to use only limited amount of arbitrage trades (proportional to the trader's information parameter $\tau$) per one short position in the option contract. $\aleph_\tau$ is also aware that that $\gamma > r$. At time t, $\aleph_\tau$ enters long (resp. short) forward contract with maturity $t + \Delta t$ when the trader knows that the stock will be "up" or "middle" (resp."down").[16]

---

[16] The payoff from this arbitrage strategy with $S_{k\Delta t;\mathbb{Q}} = x$ is

$$P_{t+\Delta t}(\aleph_\tau, arb) = x \begin{cases} (\gamma - r)\Delta t + \varrho\sqrt{\Delta t} & w.p.\ \left(\frac{\sigma^2}{2\varrho^2} + \frac{1}{\rho}(\mu - \gamma)\sqrt{\Delta t}\right) \\ (\gamma - r)\Delta t & w.p.\ \left(1 - \frac{\sigma^2}{\varrho^2}\right) \\ (r - \gamma)\Delta t + \varrho\sqrt{\Delta t} & w.p.\ \left(\frac{\sigma^2}{2\varrho^2} - \frac{1}{\rho}(\mu - \gamma)\sqrt{\Delta t}\right) \end{cases}$$

Suppose $\aleph_\tau$ takes a short position in a European option contract with payoff $f_{u,\Delta t}$ (resp. $f_{d,\Delta t}$, or $f_{d,\Delta t}$) when the stock-price is up (resp. "down" or "middle") at maturity $t + \Delta t$. Because $\aleph_\tau$ knows the stock direction. the payoff is given by $P_{t+\Delta t}(\aleph_\tau, arb, portfolio) =$

$$\begin{cases} x\Delta^{(\tau)}(1 + \gamma\Delta t + \varrho\sqrt{\Delta t}) + +xU^{(\tau,u)} - f_{u,\Delta t} & w.p.\ \left(\frac{\sigma^2}{2\varrho^2} + \frac{1}{\rho}(r - \gamma)\sqrt{\Delta t}\right) \\ x\Delta^{(\tau)}(1 + \gamma\Delta t) + xM^{(\tau,u)} - f_{m,\Delta t} & w.p.\ \left(1 - \frac{\sigma^2}{\varrho^2}\right) \\ \Delta^{(\tau)}(1 + \gamma\Delta t - \varrho\sqrt{\Delta t}) + xD^{(\tau,u)} - f_{d,\Delta t} & w.p.\ \left(\frac{\sigma^2}{2\varrho^2} - \frac{1}{\rho}(r - \gamma)\sqrt{\Delta t}\right) \end{cases}.$$

To construct a riskless portfolio $\aleph_\tau$ chooses: (i) delta position $\Delta^{(\tau)} = \frac{f_{u,\Delta t} - f_{d,\Delta t}}{2x\varrho\sqrt{\Delta t}}$ in the stock; (ii) arbitrage position $U^{(\tau,u)}$ when the trader knows that the stock will be in "up" position,

$$U^{(\tau,u)} = \phi \begin{cases} \left(\left(\frac{1}{2\varrho x} - \frac{V(\tau)^2\sigma^2}{2\varrho x\phi\varrho^2} - \frac{1}{2\varrho x}\right) + \left(-2\frac{\mu - D_\tau\tau - \gamma}{\phi 2\varrho xV(\tau)\sigma}\right)\sqrt{\Delta t}\right)f_{u,\Delta t} + \\ + \left(\left(\frac{1}{2\varrho x} - \frac{V(\tau)^2\sigma^2}{2\varrho x\phi\varrho^2} - \frac{1}{2\varrho x}\right) + 2\frac{\mu - D_\tau\tau - \gamma}{2\varrho x\phi V(\tau)\sigma}\sqrt{\Delta t}\right)f_{d,\Delta t} - \\ \left(\frac{-2}{\phi 2\varrho x}\left(1 - \frac{V(\tau)^2\sigma^2}{\varrho(\tau)^2}\right) - \frac{\varrho}{\phi}\sqrt{\Delta t} + \frac{2}{2\varrho x}\right)f_{m,\Delta t} \end{cases};$$

(iii) arbitrage position $D^{(\tau,u)}$ when the trader knows that the stock will be in "down" position,

Page | 7

*Proposition 3.* Suppose the trader $\aleph_\tau, \tau \in R$ has information about the stock-price mean return and volatility leading to a specific for $\aleph_\tau$-risk-neutral dynamics given by $S_{t;\mathbb{Q}}(\tau) = S_{t;\mathbb{Q}}^{(r,\sigma(\tau))}(\tau) = e^{\left(r - D_\tau \tau - \frac{1}{2}V(\tau)^2 \sigma^2\right)t + V(\tau)\sigma B(t)}, t \geq 0$. Then the value of $\aleph_\tau's$ short position in a European call contract with maturity price $C^{(\aleph_\tau)}\left(S_T^{(\aleph_\tau)}, T\right) = \max\left(0, S_T^{(\aleph_\tau)} - K\right)$ is given by

$$C^{(\aleph_\tau)}\left(S_t^{(\aleph_\tau)}, t\right) = e^{-K^{(\aleph_\tau)}(T-t)} S_t^{(\aleph_\tau)} N(d_1) - K e^{-r(T-t)} N(d_2) \tag{10}$$

where $K^{(\aleph_\tau)} = D_\tau \tau, \tau \in R, D_\tau \geq 0$, $d_1 = \dfrac{\ln\left(\frac{S_t^{(\aleph_\tau)}}{K}\right) + \left(r - K^{(\aleph_\tau)} + \frac{\sigma^2}{2} V(\tau)^2\right)(T-t)}{\sigma V(\tau)\sqrt{T-t}}$ and

$d_2 = d_1 - \sigma V(\tau)\sqrt{T-t}, t \in [0,T]$, $V(\tau) = A_\tau \exp(-B_\tau \tau), A_\tau \geq 0, B_\tau \geq 0$.

## 2.4. Option Pricing Formula for a Trader with Information on the True Discount Factor

We use a setting similar to that in Section 2.3. Estimating the implied risk-neutral tree with proprietary methodology, or using other information, $\aleph_\tau$ knows that the true discount factor is $D^{(r)} = \exp(-rt)$, and thus the true risk-neutral price dynamics given by $S_{t;\mathbb{Q}}^{(r,\sigma)}$ in (3). $\aleph_\tau$ is also aware that the market perceived discount factor is $D^{(R)} = \exp(-Rt)$; that is, the market perceived risk-neutral dynamics is $S_{t;\mathbb{Q}}^{(R,\sigma)} = e^{\left(R - \frac{1}{2}\sigma^2\right)t + \sigma B(t)}, t \geq 0$. While the trader knows the true discount factor rate, $\aleph_\tau$ can only borrow and lend at rate R. The next lemma follows directly from Lemma 2.

---

$$D^{(\tau,u)} = \phi \left\{ \begin{array}{l} \left(\left(\frac{1}{2\varrho x} - \frac{V(\tau)^2 \sigma^2}{2\varrho x \phi \varrho^2} - \frac{1}{2\varrho x}\right) + \left(-2 \frac{\mu - D_\tau \tau - \gamma}{\phi 2\varrho x V(\tau)\sigma}\right)\sqrt{\Delta t}\right) f_{u,\Delta t} + \\ + \left(\left(\frac{1}{2\varrho x} - \frac{V(\tau)^2 \sigma^2}{2\varrho x \phi \varrho^2} - \frac{1}{2\varrho x}\right) + 2 \frac{\mu - D_\tau \tau - \gamma}{2\varrho x \phi V(\tau)\sigma}\sqrt{\Delta t}\right) f_{d,\Delta t} - \\ \left(\frac{-2}{\phi 2\varrho x}\left(1 - \frac{V(\tau)^2 \sigma^2}{\varrho(\tau)^2}\right) - \frac{\varrho}{\phi}\sqrt{\Delta t} + \frac{2}{2\varrho x}\right) f_{m,\Delta t} \end{array} \right\};$$

(iv) arbitrate position $M^{(\tau,u)}$ when the knows that the stock will be in "middle "position

$$M^{(\tau,u)} = \left\{ \begin{array}{l} \left(\left(\frac{1}{2\varrho x} - \frac{V(\tau)^2 \sigma^2}{2\varrho x \phi \varrho^2}\right) + \left(-2 \frac{\mu - D_\tau \tau - \nu}{\phi 2\varrho x V(\tau)\sigma}\right)\sqrt{\Delta t}\right) f_{u,\Delta t} + \\ + \left(\left(\frac{1}{2\varrho x} - \frac{V(\tau)^2 \sigma^2}{2\varrho x \phi \varrho^2}\right) + 2 \frac{\mu - D_\tau \tau - \nu}{2\varrho x \phi V(\tau)\sigma}\sqrt{\Delta t}\right) f_{d,\Delta t} - \\ \left(\frac{-2}{\phi 2\varrho x}\left(1 - \frac{V(\tau)^2 \sigma^2}{\varrho(\tau)^2}\right) - \frac{\varrho}{\phi}\sqrt{\Delta t}\right) f_{m,\Delta t} \end{array} \right\},$$

where (iii) $K^{(\aleph_\tau)} := D_\tau \tau, \tau \in R, D_\tau \geq 0$ is the dividend rate. $\aleph_\tau$ enjoys due to clairvoyance, and (iv) $V(\tau) = A_\tau \exp(-B_\tau \tau), A_\tau \geq 0, B_\tau \geq 0$ is the parameter of informed volatility due to $\aleph_\tau's$ knowledge about the true Sharpe ratio $\theta = \frac{\mu - r}{\sigma}$. Then $\aleph_\tau$ obtains informed risk-neutral stock dynamics $S_{t;\mathbb{Q}}(\tau) = S_{t;\mathbb{Q}}^{(r,\sigma(\tau))}(\tau) = e^{\left(r - D_\tau \tau - \frac{1}{2}V(\tau)^2 \sigma^2\right)t + V(\tau)\sigma B(t)}, t \geq 0$.

*Lemma 4. Let (2) and (3) hold and $R > 0$ is fixed. Then the binomial tree*

$$S_{(k+1)\Delta t;\mathbb{P}} := S_{k\Delta t;\mathbb{P}} \begin{cases} 1 + \lambda\Delta t + \sigma\sqrt{\Delta t} & w.p.\ p_{R,\Delta t} = \frac{1}{2} + \frac{\mu-\lambda}{2\sigma}\sqrt{\Delta t}, \\ 1 + \lambda\Delta t - \sigma\sqrt{\Delta t} & w.p.\ 1 - p_{R,\Delta t}, \lambda := r - R + \frac{\sigma^2}{2}, \end{cases}$$

$k := 0, \dots, n-1, n\Delta t = T$ is approximating $S_{t;\mathbb{P}}, t \in [0, T]$ in $\Rightarrow^W$-sense. The corresponding risk-neutral tree approximating $S_{t;\mathbb{Q}}$ in the sense of $\Rightarrow^W$ is given by

$$S_{(k+1)\Delta t;\mathbb{Q}} := S_{k\Delta t;\mathbb{Q}} \begin{cases} 1 + \lambda\Delta t + \sigma\sqrt{\Delta t} & w.p.\ q_{v,\Delta t}, \\ 1 + \lambda\Delta t - \sigma\sqrt{\Delta t} & w.p.\ 1 - q_{v,\Delta t} \end{cases}$$

*where* $q_{v,\Delta t} = \frac{1}{2} + \frac{R-\frac{\sigma^2}{2}}{2\sigma}\sqrt{\Delta t}$ *is the risk-neutral probability for upward stock movement.*

According to Lemma 4 $\aleph_\tau$ can enter forward contract initiated at t, with arbitrage opportunity and forward price $\mathbb{F}_{t;M} = x(1 + R\Delta t), x = S_{t;\mathbb{P}}$, and terminal time $t + \Delta t$. The arbitrage-trade payoff (under $\mathbb{Q}$) of the long forward contract is $S_{(k+1)\Delta t;\mathbb{Q}} - \mathbb{F}_{t;M}$ with a present value of $x(r - R)\Delta t$. Per one short position in the option contract, $\aleph_\tau$ is permitted to use only a limited amount of arbitrage trades (proportional to the trader's information parameter $\tau$). $\aleph_\tau$ takes a short position in a European option contract with payoff $f_{u,\Delta t}$ (resp. $f_{d,\Delta t}$) when the stock price is "up" (resp. "down") at maturity $t + \Delta t$. The trader also takes an arbitrage-long forward contract (resp. arb-short forward) with payoff $x(r - R)\Delta t$ (resp. $x(R - r)\Delta t$) when $R < r$ (resp. $R \geq r$)[17].

*Proposition 4. Suppose the trader $\aleph_\tau, \tau \in R$ has information about the true discount factor*

---

[17] To hedge the risk associated with this position, $\aleph_\tau$ takes (i)$\Delta^{(\tau)}$ —position in the underlying stock to hedge the "up" and "down" risk; (ii) $D_\tau$ — position if $R < r$ (resp. $-D_\tau$, if $R \geq r$) in an arbitrage trade. Then, the payoff of $\aleph_\tau$ at $t + \Delta t$, after closing all positions, is $P(\aleph_\gamma; t + \Delta t) =$

$$\begin{cases} x\Delta^{(\tau)}(1 + \lambda\Delta t + \sigma\sqrt{\Delta t}) + D_\tau(r - R)\Delta t - f_{u,\Delta t} & w.p.\ p_{R,\Delta t} = \frac{1}{2} + \frac{\mu-\lambda}{2\sigma}\sqrt{\Delta t}, \\ x\Delta^{(\tau)}(1 + \lambda\Delta t - \sigma\sqrt{\Delta t}) + D_\tau(r - R)\Delta t - f_{d,\Delta t} & w.p.\ 1 - p_{R,\Delta t}, \lambda := r - R + \frac{\sigma^2}{2} \end{cases}$$

The limitation on arbitrage trade is defined by the parameter $C^{(\tau)}\tau, C^{(\tau)} \geq 0, \tau \in R$, and let $D_\tau = \frac{R+C^{(\tau)}\tau}{r-R}f_{0,t,\Delta t}^{(R)}$, where $f_{0,t,\Delta t}^{(R)} = (1 - R\Delta t)(q_{v,\Delta t}f_{u,\Delta t} + (1 - q_{v,\Delta t})f_{d,\Delta t})$ is the market perceived binomial option value at t.

Requiring equality of payoffs' outcomes and discounting the obtained riskless portfolio with $D^{(r)}$ leads to the following formula for $\aleph_\tau'$s binomial option value of t: $f_{0,t}^{(\tau)} = e^{-R^{(\tau)}}(q_{v,\Delta t}^{(\tau)}f_{u,\Delta t} + (1 - q_{v,\Delta t}^{(\tau)})f_{d,\Delta t})$, where $q_{v,\Delta t}^{(\tau)} := \frac{1}{2} + \frac{R-\frac{\sigma^2}{2}}{2\sigma}\sqrt{\Delta t}$ with $R^{(\tau)} := R + C^{(\tau)}\tau$. The corresponding $\aleph_\tau'$s specific risk-neutral dynamics is given by

$$dS_t^{(\aleph_\tau)} = R^{(\tau)}S_t^{(\aleph_\tau)}dt + \sigma S_t^{(\aleph_\tau)}dB(t).$$

*implying that the $\aleph'_\tau s$ specific risk-neutral dynamics is given by* $dS_t^{(\aleph_\tau)} = R^{(\tau)} S_t^{(\aleph_\tau)} dt + \sigma S_t^{(\aleph_\tau)} dB(t)$. *Then the value of* $\aleph_\tau's$ *short position in a European call contract with maturity price* $C^{(\aleph_\tau)}\left(S_T^{(\aleph_\tau)}, T\right) = max\left(0, S_T^{(\aleph_\tau)} - K\right)$ *is given by*

$$C^{(\aleph_\tau)}\left(S_t^{(\aleph_\tau)}, t\right) = e^{-C^{(\tau)}\tau(T-t)} S_t^{(\aleph_\tau)} N(d_1) - K e^{-R^{(\tau)}(T-t)} N(d_2), \qquad (11)$$

*where* $R^{(\tau)} := R + C^{(\tau)}\tau, \tau \in R, C_\tau \geq 0$, $d_1 = \dfrac{\ln\left(\frac{S_t^{(\aleph_\tau)}}{K}\right) + \left(R + \frac{\sigma^2}{2}\right)(T-t)}{\sigma\sqrt{T-t}}$, *and* $d_2 = d_1 - \sigma\sqrt{T-t}$

## 2.5 Option Pricing in the Mean-Variance Framework

Suppose the short-position-option-holder, $\aleph_\tau$, is facing a hedge-turnover constraint (per one share traded) of $HT_{\Delta t} = \varphi^2 n, \varphi > 0, n = \dfrac{T}{\Delta t}, T$ is the option maturity time. $\aleph$ would like to keep $HT_{\Delta t}$ limited to a bound of $HT_{\Delta t} \leq B^2$ for some fixed constant $B > 0$. Thus, $\Delta t = \dfrac{T}{n} = \dfrac{T\varphi^2}{HT_{\Delta t}} \geq \dfrac{T\varphi^2}{B^2}$ [18]. Then the risk-adjusted process is

$$\begin{cases} \Delta^{(p_{\Delta t})} u_{\Delta t} - f_{u,\Delta t} = \Delta^{(p_{\Delta t})} y \left(1 + \gamma\Delta t + \sqrt{\dfrac{1-p_{\Delta t}}{p_{\Delta t}}} \sigma\sqrt{\Delta t}\right) - f_{u,\Delta t} \text{ w. p. } p_{\Delta t} \\ \Delta^{(p_{\Delta t})} d_{\Delta t} - f_{d,\Delta t} = \Delta^{(p_{\Delta t})} y \left(1 + \gamma\Delta t - \sqrt{\dfrac{p_{\Delta t}}{1-p_{\Delta t}}} \sigma\sqrt{\Delta t}\right) - f_{d,\Delta t} \text{ w. p. } 1 - p_{\Delta t} \end{cases}$$

---

[18] Because of the limitation to hedge in continuous time, $\aleph$ opts to optimize the mean-variance tradeoff of the hedged portfolio at $t + \Delta t, t = 0, \Delta t, \ldots, (n-1)\Delta t$: $P^{(t+\Delta t)} :=$

Here, $y = x(1 + HT_{\Delta t}^2)$, where $x = S_t$, and the derivative-price dynamics are given by the tree $f_{(k+1)\Delta t} :=$
$\begin{cases} f_{k\Delta t}^+ = f_{u,\Delta t} \text{ w. p. } p_{\Delta t}, \\ f_{k\Delta t}^- = f_{d,\Delta t} \text{ w. p. w. p. } 1 - p_{\Delta t} \end{cases}$. $\aleph'_\tau s$ goal is to maximize the objective function $\mathbb{E}_t P^{(t+\Delta t)}$ given that $var_t P^{(t+\Delta t)} \leq \varepsilon_{\Delta t}^2$, where $\varepsilon_{\Delta t} := \lambda(f_{u,\Delta t} - f_{d,\Delta t})\sqrt{p_{\Delta t}(1 - p_{\Delta t})}$ with, $\lambda = \lambda_0 + \dfrac{\varphi}{B}, \lambda_0 > 0$. Because

$var_t P^{(t+\Delta t)} = p_{\Delta t}(1 - p_{\Delta t})\left(\Delta^{(p_{\Delta t})} y \left(\dfrac{1}{\sqrt{p_{\Delta t}(1-p_{\Delta t})}}\sigma\sqrt{\Delta t}\right) - (f_{u,\Delta t} - f_{d,\Delta t})\right)^2 \leq \varepsilon_{\Delta t}^2$, the optimal delta position is $\Delta^{(p_{\Delta t})(*)} y = \dfrac{\varepsilon}{\sigma\sqrt{\Delta t}} +$
$\dfrac{(f_{u,\Delta t}-f_{d,\Delta t})}{\sigma\sqrt{\Delta t}}\sqrt{p_{\Delta t}(1 - p_{\Delta t})}$, and thus the optimal conditional portfolio mean is given by $\mathbb{E}_t P^{(t+\Delta t)(*)} =$
$\left\{\dfrac{\varepsilon}{\sigma\sqrt{\Delta t}} + \dfrac{(f_{u,\Delta t}-f_{d,\Delta t})}{\sigma\sqrt{\Delta t}}\sqrt{p_{\Delta t}(1 - p_{\Delta t})}\right\}(1 + \gamma\Delta t) - p_{\Delta t}f_{u,\Delta t} - (1 - p_{\Delta t})f_{d,\Delta t}$. This leads to the option price at the nodes of the tree of the form $f_t = e^{-r\Delta t}(Q^{(\Delta t,\lambda)} f_{u,\Delta t} + (1 - Q^{(\Delta t,\lambda)}) f_{d,\Delta t})$ where the "risk-adjusted-probabilities" are $Q^{(\Delta t,\lambda)} = p_{\Delta t} - \theta(1 + \lambda)\sqrt{p_{\Delta t}(1 - p_{\Delta t})}\sqrt{\Delta t}$ and $1 - Q^{(\Delta t,\lambda)}$. This allows us to value the option on the "risk-adjusted tree" $S_{(k+1)\Delta t} := S_{k\Delta t}\begin{cases} 1 + \gamma\Delta t + \sqrt{\dfrac{1-p_{\Delta t}}{p_{\Delta t}}}\sigma\sqrt{\Delta t} \text{ w. p. } Q^{(\Delta t,\lambda)}, \\ 1 + \gamma\Delta t - \sqrt{\dfrac{p_{\Delta t}}{1-p_{\Delta t}}}\sigma\sqrt{\Delta t} \text{ w. p. } 1 - Q^{(\Delta t,\lambda)} \end{cases}$ $k := 0, \ldots, n - 1, n\Delta t = T$ with optimal $\Delta t = \dfrac{T\varphi^2}{B^2}$. Next, to have the limit of the pricing tree as $\Delta t \downarrow 0$, we should have either either $\varphi = 0$, or $B = \infty$. In both cases, $\lambda = \lambda_0$.

Page | 10

$$S_t = S_t^{(r+\lambda\sigma,\sigma)} = e^{\left(r-(\gamma-r)\lambda_0 - \frac{1}{2}\sigma^2\right)t + \sigma B(t)}, t \geq 0 \qquad (12)$$

That is, we can view the option contract as written against the stock with dividend rate $(\gamma - r)\lambda_0$. For a call option with strike K and maturity T, the risk-adjusted dynamics (12) imply the formula: $C(S_t, t) = e^{-(\gamma-r)\lambda_0(T-t)} S_t N(d_1) - K e^{-r(T-t)} N(d_2)$, where $d_1 = \frac{\ln\left(\frac{S_t}{K}\right) + \left(r-(\gamma-r)\lambda_0 + \frac{\sigma^2}{2}\right)(T-t)}{\sigma\sqrt{T-t}}$, $d_2 = d_1 - \sigma\sqrt{T-t}$ and N is the cumulative standard normal distribution function.

## 3. MARKETS WITH LIMITED ARBITRAGE OPPORTUNITIES DRIVEN BY LOG-ROSENBLATT-HERMITE TYPE PROCESSES

Grossman and Stiglitz (1980) argued that it is practically impossible to take advantage of all the arbitrage opportunities at every instant of time when information is costly, since there must be some benefit to the informed investor to release this information to uninformed investors at no charge. Deriving an optimal investment policy of a risk-averse investor in a market with arbitrage opportunities, Liu and Longstaff (2004) find that it is often optimal to underinvest in the arbitrage by taking a smaller position than collateral constraints allow. The possibility of discrepancies leading to arbitrage may arise in real markets (e.g., see Dwyer et al., 1996) so it is desirable to study trading strategies which produce arbitrage opportunities (see Salopek, 1998).There are no guarantees, however, that in the real world there will be convergence of frictionless equilibrium price model values to their observed prices. Since observed markets seem to possess memory[19], fractional Brownian motion [20] and Rosenblatt processes [21] may be useful for modeling observed prices as opposed to frictionless equilibrium prices[22]. The goal of this section is to show that empirical evidence confirms the validity of pricing model exhibiting long-range dependence and limited arbitrage opportunities.

### 3.1. The Binary Tree Model and Limited Arbitrage

In this section we shall extend the KSRF- binomial tree price model to a binary pricing model which will be more suited as a starting point in developing binary models with long–range dependence. Starting with a binary tree model with fixed frequency of stock movements, we assume that the trader, designated as ℶ, seeking access to limited arbitrage opportunities can explore arbitrage trading activities. ℶ has collected intraday high-frequency stock data $S_{t,obs}$, $t > 0$ in different frequency $\Delta t > 0$, $\mathfrak{N}\Delta t = \mathcal{T}$, where $\mathcal{T}$ is the sample window and $\mathfrak{N}$ is the sample size. . Let T be ℶ's investing horizon and $T = \mathcal{N}\Delta t$. The data ℶ has collected allows the trader to have estimates $\hat{\mu}, \hat{\sigma}, \hat{p}_{\Delta t} = \hat{g} + \hat{v}\sqrt{\Delta t}$, for the parameters of the following discrete binary price-process dynamics:

$$S_{(k+1)\Delta t} = S_{k\Delta t}\left(1 + \mu\Delta t + \left(\frac{1-p_{\Delta t}}{p_{\Delta t}}\right)^{\frac{\xi_{k+1,\mathcal{N}}}{2}} \sigma\sqrt{\Delta t}\xi_{k+1;\mathcal{N}}\right) \tag{13}$$

---

[19] See, for example, Lo (1991), Campbell, et al. (1997), Andersson (1998). Diebold (2001) Nielsen (2010), Johansen ( 2011), and Caporale, Gil-Alana (2014)
[20] See Bayraktar et al. (2004), Rostek (2009) and the references there in
[21] See Taqqu (1975) Rosenblatt (1985),and Taqqu (2011)
[22] See Goldenberg (1986). Soros (1994, Chapter 1, Part 3) argues: "Returning to economic theory, it can be argued that it is the participants' bias that renders the equilibrium position unattainable. The target toward which the adjustment process leads incorporates a bias, and the bias may shift in the process. When that happens, the process aims not at an equilibrium but at a moving target. To put matters into perspective, we may classify events into two categories: humdrum, everyday events that are correctly anticipated by the participants and do not provoke a change in their perceptions, and unique, historical events that affect the participants' bias and lead to further changes. The first kind of event is susceptible to equilibrium analysis, the second is not: it can be understood only as part of a historical process."

The parameters of the binary tree model (13) are (i) $\mu \in \mathbb{R}$ – stock instantaneous mean return, and; (ii) $\sigma > 0$ – stock volatility; (iii) $p_{\Delta t} = g + v\sqrt{\Delta t}$, stock probability for non-negative return in $\Delta t$, for some fixed $g \in (0,1), v \in \mathbb{R}$, and $\Delta t > 0$ is sufficiently small for $p_{\Delta t} \in (0,1)$; (iv) for every $\mathcal{N} = 2, \ldots, \mathbb{N}, \{\xi_{k,\mathcal{N}}, k = 1,2, \ldots\}$ is a sequence of iid random signs $\mathbb{P}(\xi_{k,\mathcal{N}} = 1) =$

$1 - \mathbb{P}(\xi_{k,\mathcal{N}} = -1) = p_{\Delta t}$. Then for every fixed set of parameters $(\mu, \sigma, g, v) \in \mathbb{R} \times (0, \infty) \times (0,1) \times \mathbb{R}$, the discrete price process generated by (2.1) converges in $\Rightarrow^w$ to the GBM[23]

$$S_t = S_t^{(\mu,\sigma^2)} = S_0 e^{\left(\mu - \frac{1}{2}\sigma^2\right)t + \sigma B(t)}, t > 0. \tag{14}$$

We assume that price processes (13) and (14) are defined on the natural world, designated by a stochastic basis $(\Omega, (\mathcal{F}_t, t \in [0,T]), \mathbb{P})$. The generated risk-neutral binary price process defined on the equivalent stochastic basis $(\Omega, (\mathcal{F}_t, t \in [0,T]), \mathbb{Q})$ is given by [24]

$$S_{(k+1)\Delta t} = S_{k\Delta t}\left(1 + \mu\Delta t + \left(\frac{1-p_{\Delta t}}{p_{\Delta t}}\right)^{\frac{\xi_{k+1,N}}{2}} \sigma\sqrt{\Delta t}\eta_{k+1;\mathcal{N}}\right) \tag{15}$$

where for every $\mathcal{N} = 2, \ldots, \mathbb{N}, \{\eta_{k,N}, k = 1,2, \ldots\}$ is a sequence of independent and identically distributed (iid) random signs $\mathbb{Q}(\eta_{k,\mathcal{N}} = 1) = 1 - \mathbb{Q}(\eta_{k,\mathcal{N}} = -1) = q_{\Delta t}$ where

$$q_{\Delta t} = p_{\Delta t} - \sqrt{p_{\Delta t}(1 - p_{\Delta t})}\theta\sqrt{\Delta t}, \theta = \frac{\mu - r}{\sigma} \tag{16}$$

The $C[0,1]$ – random broken line generated by the tree (15) converges to the risk-neutral GBM:

$$S_t = S_t^{(r,\sigma^2)} = S_0 e^{\left(r - \frac{1}{2}\sigma^2\right)t + \sigma B(t)}, t > 0. \tag{17}$$

**3.2. The Binary Tree Model with Random Frequency of Stock Movements**

---

[23] See Lemma 1 in Kim et al (2016a), where the corresponding weak convergence in C[0,1] is shown as well. See also Davidov and Rotar (2008).
[24] The proof is almost identical to that in Lemma 1, Kim et al (2016).

Consider in the tree (13) subordinated time steps $\Delta^{(k+1)} := \tau^{(k+1)} - \tau^{(k)}, \tau^{(k)} := \tau(k\Delta t)$, where $\tau(t)$ is a non-decreasing Lévy process (a Lévy- subordinator [25]) with $\tau(0) = 0$ on stochastic basis $\left(\Omega^{(\tau)}, \left(\mathcal{F}_t^{(\tau)}, t > 0\right), \mathbb{P}^{(\tau)}\right)$, and is independent of the sequence $\{\xi_{k,\mathcal{N}}, k = 0,1,...\}$. Consider the subordinated tree model

$$S_{\tau^{(k+1)}} = S_{\tau^{(k)}}\left(1 + r\Delta t + \rho\Delta^{(k+1)} + \left(\frac{1-p_{\Delta t}}{p_{\Delta t}}\right)^{\frac{\xi_{k+1,\mathcal{N}}}{2}} \sigma\sqrt{\Delta^{(k+1)}}\xi_{k+1;\mathcal{N}}\right) \quad (18)$$

Then the corresponding discrete price process weakly converges in $D[0,1]$ to

$$S_t^{(\tau)} = S_0 e^{rt + \rho\tau(t) + \sigma B(\tau(t))}, t > 0 \text{ [26]} \quad (19)$$

where $B(t)$ is a Brownian motion on $(\Omega, (\mathcal{F}_t, t \in [0,T]), \mathbb{P})$ and is independent of $\{\tau(t), t \geq 0\}$[27]. Furthermore, the martingale measure $\mathbb{Q}$ is defined by the Radon-Nikodym derivative $\Psi(t) = \exp\left\{\psi B(t) - \frac{1}{2}\psi^2 t\right\}, t \geq 0$, where $\psi = -\frac{\rho + \frac{\sigma^2}{2}}{\sigma}$ is the market price for risk, see Hurst et al. (1999). Consider a European call option with exercise price K and time to maturity T. Let $K^{(r,T,t)} = Ke^{-r(T-t)}$ be the discounted exercise price. Then the option value $C_t$ at $t \in [0,T)$ is given by

$$C_t = S_t F^{(+)}\left(\log\left(\frac{S_t}{K^{(r,T,t)}}\right)\right) - K^{(r,T,t)} F^{(+)}\left(\log\left(\frac{S_t}{K^{(r,T,t)}}\right)\right), \quad (20)$$

where $F^{(+)}(x) = \int_0^\infty F_{N(0,1)}\left(\frac{x + \frac{y}{2}}{\sqrt{y}}\right) dF_{Y_t^{(T)}}(y)$, $F_{N(0,1)}$ is the standard normal distribution function and $F_{Y_t^{(T)}}$ is the distribution function of $Y_t^{(T)} = \sigma^2(\tau(T) - \tau(t))$.

In particular, if for $\alpha \in (1,2)$,

$$Y_t^{(T)} = C_\alpha^{(T-t)} V_\alpha \quad (21)$$

where $C_\alpha^{(T-t)} = 2\sigma^2 \cos\left(\frac{\pi\alpha}{4}\right)^{\frac{2}{\alpha}} (T-t)^{\frac{2}{\alpha}}$ and $V_\alpha \sim S_{\frac{\alpha}{2}}(1,1,0)$ is a standard $\frac{\alpha}{2}$-stable subordinator (see Samorodnitsky and Taqqu, (1994), that is, an $\frac{\alpha}{2}$-stable random variable with unit skewness and scale parameters.[28] Then (8) is the value of A European call when the pricing process is driven by α-stable process (see Hurst et al., 1999).

---

[25] See Rachev et al. (2011), Bianchi(2010), and Kim et al.(2010).
[26] Note that because the parameter $\rho$ is strictly positive, and the distribution of $\tau(1)$ can be quite general, the discrete and continuous models ,(3.16) and (3.17) are flexible enough to be used as models for the price process on the natural world $(\Omega, (\mathcal{F}_t, t \in [0,T]), \mathbb{P})$. The models are general enough to compensate for the choice of the mean instantaneous deterministic drift to be equal to the risk free rate r.
[27] The omitted proof is similar to the one in Karandikar and Rachev (1999)
[28] $\tau(t), t \geq 0$ is a $Lévy - stable\ subrodinator$, if it is a $Lévy$-process (that is a process with independent and stationary increments, starting at 0) and having an unit increment given by $C_\alpha^{(1)}$.

## 3.3. The Binary Tree Model with Limited Arbitrage

Suppose at each instant $k\Delta t$, $k = 0,1,\ldots$, ב knows (with probability 1) the stock direction at $(k+1)\Delta t$. Knowing that the price will be "up" (resp. "down")[29] at $(k+1)\Delta t$, ב enters a long-forward (resp. short-forward) contract with an un-informed trader.[30] Applying the tree model, ב's payoff (at time $t + \Delta t$,) is

$$P_{t+\Delta t}(\aleph_\tau) = x \begin{cases} u_{\Delta t} - e^{r\Delta t} = \rho\Delta^{(k+1)} + \sqrt{\frac{1-p_{\Delta t}}{p_{\Delta t}}}\sigma\sqrt{\Delta^{(k+1)}} & \text{w.p. } p_{\Delta t} \\ d_{\Delta t} - e^{r\Delta t} = \rho\Delta^{(k+1)} - \sqrt{\frac{p_{\Delta t}}{1-p_{\Delta t}}}\sigma\sqrt{\Delta^{(k+1)}} & \text{w.p. } 1 - p_{\Delta t} \end{cases} \quad (22)$$

Suppose ב takes a short position in a European option contract with payoff $f_{u,\Delta t}^{(t+\Delta t)}$ (resp. $f_{d,\Delta t}^{(t+\Delta t)}$) when the stock-price is up (resp. down), at maturity $t + \Delta t$[31]. Consider the arbitrage process

$$S_{\tau^{(k+1)}}^{(A)} = S_{\tau^{(k)}}^{(A)}\left(1 + r\Delta t + \rho\Delta^{(k+1)} + \left(\frac{1-p_{\Delta t}}{p_{\Delta t}}\right)^{\frac{\xi_{k+1,\mathcal{N}}}{2}}\sigma\sqrt{\Delta^{(k+1)}}\eta_{k+1;\mathcal{N}}^{(a)}\right),$$

where $\eta_{k,\mathcal{N}}^{(a)}$ are iid random signs with $\mathbb{P}(\eta_{k,\mathcal{N}}^{(a)} = 1) = 1 - \mathbb{P}(\eta_{k,\mathcal{N}}^{(a)} = -1) = q^{(a)}$. Then the conditional mean is given by $\mathbb{E}_{\tau^{(k)}}^{(\mathbb{Q})}\frac{S_{\tau^{(k+1)}}^{(A)}}{S_{\tau^{(k)}}^{(A)}} = 1 + r\Delta t + 2\rho p_{\Delta t}\Delta^{(k+1)}$, while $\text{Var}_{\tau^{(k)}}^{(\mathbb{Q})}\frac{S_{\tau^{(k+1)}}^{(\mathbb{Q})}}{S_{\tau^{(k)}}^{(\mathbb{Q})}} = 0$;

---

[29] That is, $u_{\Delta t} = x\left(1 + r\Delta t + \rho\Delta^{(k+1)} + \sqrt{\frac{1-p_{\Delta t}}{p_{\Delta t}}}\sigma\sqrt{\Delta^{(k+1)}}\right)$, $x = S_{\tau^{(k)}}$, (resp. $d_{\Delta t} = x\left(1 + r\Delta t + \rho\Delta^{(k+1)} - \sqrt{\frac{p_{\Delta t}}{1-p_{\Delta t}}}\sigma\sqrt{\Delta^{(k+1)}}\right)$

[30] The forward price is $xe^{r\Delta t} = x(1 + r\Delta t)$ with the contract's maturity $\Delta t$.

[31] ב uses $\Delta^{(a)}$ arbitrage- forward contracts to hedge the stock the "up" and "down" risk, as it costs the trader nothing to

enter those forward contracts. ב's arbitrage portfolio $t + \Delta t$. ב –position at $t + \Delta t$ in the arbitrage portfolio is

$$P(ב; t + \Delta t) = \begin{cases} \Delta^{(a)}(u_{\Delta t} - e^{r\Delta t}) - f_{u,\Delta t}^{(t+\Delta t)} & \text{w.p. } p_{\Delta t} \\ \Delta^{(a)}(e^{r\Delta t} - d_{\Delta t}) - f_{d,\Delta t}^{(t+\Delta t)} & \text{w.p. } 1 - p_{\Delta t} \end{cases}$$

Equating both outcomes leads to $\Delta^{(a)} = \frac{f_{u,\Delta t}^{(t+\Delta t)} - f_{d,\Delta t}^{(t+\Delta t)}}{2\rho\Delta^{(k+1)} + \frac{1}{\sqrt{p_{\Delta t}(1-p_{\Delta t})}}\sigma\sqrt{\Delta^{(k+1)}}}$. The riskless portfolio's value is $\Delta^{(a)}(u_{\Delta t} - e^{r\Delta t}) - f_{u,\Delta t}^{(t+\Delta t)}$, and thus for the derivative value at $t = k\Delta t$ we obtain $f_{\Delta t}^{(t)} = e^{-r\Delta t}\left(q^{(a)}f_{u,\Delta t}^{(t+\Delta t)} + (1 - q^{(a)})f_{d,\Delta t}^{(t+\Delta t)}\right)$, where the $q^{(a)}$ –arbitrage-probability is

$$q^{(a)} := \frac{1}{2} + \frac{p_{\Delta t} - \frac{1}{2}}{1 + \frac{2\rho}{\sigma}\sqrt{\Delta^{(k+1)}}\sqrt{p_{\Delta t}(1-p_{\Delta t})}}.$$

that is, $\frac{S^{(A)}_{\tau^{(k+1)}}}{S^{(A)}_{\tau^{(k)}}} - 1 = r\Delta t + 2\rho p_{\Delta t}\Delta^{(k+1)}, k = 1,2,\ldots$ is a sequence of random arbitrage returns.

In the case of a $\frac{\alpha}{2}$-stable subordinator (see (2.9)), $\Delta^{(k+1)}$ is distributed as $\Delta^{(k+1)} = C_\alpha^{(\Delta t)}V_\alpha$. Thus, if the log-returns in random frequency of $C_\alpha^{(\Delta t)}V_\alpha$ behaves like $r\Delta t + 2\rho p_{\Delta t}C_\alpha^{(\Delta t)}V_\alpha$ with very high probability, that is

$$\log(\frac{S^{(A)}_{\tau^{(k+1)}}}{S^{(A)}_{\tau^{(k)}}}) \sim \log(\frac{1}{S^{(A)}_0}S^{(A)}_{C_\alpha^{(\Delta t)}}) \sim r\Delta t + 2\rho p_{\Delta t}C_\alpha^{(\Delta t)}V_\alpha \quad \text{(with probability close to 1)} \quad (23)$$

then (23) could be a potential evidence of market exhibiting arbitrage opportunities.